# Breaking Down the Scoring: Interrater Reliability and National Bias in Olympic Breaking

Patrick Alexander Braeunig[1]

Saarland University

October 31, 2025

## Abstract

**Introduction:** The inclusion of Breaking in the 2024 Paris Olympic Games introduced a distinctive competition format in which two athletes compete head-to-head in battle rounds. Unlike other aesthetic sports, where athletes receive independent scores for their performances, Breaking judges assess the relative performance quality between two competitors across five predefined evaluation aspects. This may considerably increase their cognitive load and warrants a thorough examination of judging reliability and potential national bias, the tendency of judges to favor athletes from their own country.

**Methods:** Official scoring data from the 2024 Olympic Breaking events were analyzed. Interrater reliability was assessed using the Intraclass Correlation Coefficient, while national bias was estimated using mixed-effects modelling based on judges' individual scores.

**Results:** The analyses revealed low to moderate interrater reliability in Breaking, notably lower than levels reported in other aesthetic sports. Moreover, substantial national bias was detected, with judges displaying a systematic tendency to favor athletes from their own country.

**Discussion:** Although the observed national bias, despite its notable magnitude, likely had only a limited effect on final competition outcomes, the relatively low interrater reliability highlights potential weaknesses in the current scoring framework that may warrant refinement. Nonetheless, Breaking's unique system of aggregating judges' scores into discrete votes appears to reduce the influence of individual judges on competition outcomes, thereby enhancing the system's robustness against unilateral score manipulation.

---

[1]Email: patrick.braeunig@uni-saarland.de

# 1 Introduction

The inclusion of Breaking in the 2024 Paris Olympic Games marked a significant milestone in the sport's global recognition, following its first appearance on the global stage during the closing ceremony of the 1984 Los Angeles Olympics (Banes, 2004; Stevens, 2006). Now governed by the World DanceSport Federation (WDSF), Breaking has developed into a formally organized and internationally recognized competitive discipline, featuring prominent events such as Battle of the Year, Red Bull BC One, and World B-Boy Classic (Li & Vexler, 2019; Quittelier, 2015). However, its recent Olympic debut has sparked considerable debate within the Breaking community. While many practitioners welcome the sport's global visibility, others have expressed concern that the imposition of standardized judging frameworks undermines Breaking's roots as an expressive art form (Li & Vexler, 2019; Shapiro, 2004).

Emerging from the Bronx, New York, in the 1970s as a core element of hip-hop culture (Dalecki, 2011; Holman, 2004; Langnes & Fasting, 2016; Li & Vexler, 2019; Stevens, 2006; Wei et al., 2022), Breaking is widely regarded as "the most representative and well-known discipline in hip-hop dance" (Quittelier, 2015, p. 3). Rooted in American vernacular street dance, it originated and evolved in community spaces such as parks, schoolyards, and subway stations (Banes, 2004; Dalecki, 2011; Shapiro, 2004; Stevens, 2006). Beyond its artistic dimension, Breaking served as a vehicle for creative expression, identity construction, and social mobility—particularly within marginalized communities (Guarato, 2021; Langnes & Fasting, 2016; Shapiro, 2004; Wei et al., 2022). Its global dissemination was propelled by (international) practitioner networks, media exposure, and digital platforms such as YouTube (Banes, 2004; Langnes & Fasting, 2016; Shapiro, 2004; Stevens, 2006).

Traditionally, Breaking adopted a battle format, in which individuals or crews perform alternately in direct confrontation (Dalecki, 2011; Holman, 2004; Langnes & Fasting, 2016; Stevens, 2006), vying for approval. Hence, audience reactions—typically expressed through cheering—have historically shaped both the expressive features and the competitive dynamics of the sport (Guarato, 2021; Osumare, 2002; Sato, 2022). The Olympic competition format retained these defining features, including head-to-head battles and live DJ accompaniment, while introducing a more structured competition framework comprising round-robin group stages followed by knockout rounds (Dalecki, 2011; Paris Organising Committee for the 2024 Olympic and Paralympic Games, 2024; Quittelier, 2015; Shapiro, 2004; Stevens, 2006).

In Olympic Breaking, two athletes perform consecutively within each battle. A panel of expert judges evaluates their performances in direct comparison, assigning scores that reflect perceived differences in performance quality across five distinct performance quality aspects (Paris Organising Committee for the 2024 Olympic and Paralympic Games, 2024; WDSF, 2025). This comparative scoring framework distinguishes Breaking from other "aesthetic" sports (McFee, 2013, p. 2), in which athletes perform and are evaluated individually. Breaking judges must therefore assess multiple performance aspects while processing two consecutive performances and retaining detailed evaluations of the first until the second concludes—introducing unique cognitive demands.

Like other aesthetic sports, Breaking is inherently susceptible to subjectivity in performance evaluation, making it vulnerable to both random and systematic judging errors (Plessner, 1997; Plessner & Haar, 2006). Such errors may undermine the integrity of competition outcomes, erode public trust, and diminish the sport's entertainment and marketing value (Daumann, 2015; Emrich & Pierdzioch, 2015). However, Breaking's distinctive competition format and scoring system may amplify random errors and biases. The distinct multitasking and memory demands placed on judges introduce a high cognitive load, potentially reducing judging reliability (Ste-Marie, 2000). Furthermore, Breaking's head-to-head battle format introduces an inherently salient competitive relationship between pairs of athletes, which—similar to settings such as team competitions that emphasize national identity (Sala et al., 2007; Zitzewitz, 2006)—may evoke or magnify national bias. Consequently, Breaking's Olympic debut warrants thorough examination of the quality of judging with respect to both reliability and impartiality.

Interrater reliability, commonly assessed using the Intraclass Correlation Coefficient (ICC), appears to vary substantially across judged sports (Braeunig, 2025). High, though not perfect, reliability has been reported in Artistic Gymnastics (Bučar et al., 2011, 2013; Leskošek et al., 2010, 2012), Figure Skating (Weekley & Gier, 1989), and in experimental video-based assessments of gymnastics routines (Bučar Pajek et al., 2011). In contrast, moderate to low levels of interrater reliability have been documented in Taekwondo (Klein et al., 2014) and dance-related disciplines, including hip-hop competitions (Premelč et al., 2019; Sato, 2022; Sato & Hopper, 2021). This variability suggests that outcomes in dance sport competitions may be more strongly influenced by judges' assessment errors than in other aesthetic sports. However, the validity and comparability of prior reliability findings are often limited by methodological inconsistencies (Braeunig, 2025), such as the omission of the specific ICC model employed (Aleksić-Veljković et al., 2016; Klein et al., 2014; Leandro et al., 2017; Leskošek et al., 2010, 2012; Pavleski & Božilova, 2020; Sato & Hopper, 2021; Weekley & Gier, 1989), or by applying unsuitable ICC models to the scoring data (Bučar et al., 2011; Bučar et al., 2013; Bučar Pajek et al., 2011). These limitations underscore the need for more rigorous and transparent investigations of scoring reliability across aesthetic sports. Such research could help uncover systematic relationships between competition formats, scoring frameworks, and interrater reliability, thereby informing the development of more objective and reliable judging systems (Braeunig, 2025).

Beyond reliability concerns, aesthetic sports have faced persistent allegations of bias in judging. Among various known biases (Graf, 2010; Plessner & Haar, 2006), national bias—the tendency of judges to favor athletes from their own country—remains a particular concern (Heiniger & Mercier, 2021). Historical controversies, such as the 2004 Olympic Men's Artistic Gymnastics scandal—in which Russian Gymnastics President Leonid Arkayev described judging as controlled by a "mafia of judges" (Gymmedia International, 2004)—and the 2002 Winter Olympics Figure Skating scoring scandal involving "vote trading" (Zitzewitz, 2006, p. 69) between French and Russian judges—triggered extensive reforms in judging systems, judge selection procedures, and transparency policies (Berry, 2012; Zitzewitz, 2006, 2014).

Empirical research distinguishes between two principal forms of national bias (Heiniger & Mercier, 2021). The first, "patriotic bias" (Sala et al., 2007, p. 18) refers to judges awarding inflated scores to athletes from their own country (Sala et al., 2007). The second, "competitor bias" (Braeunig, 2024, p. 254), describes the tendency of judges to lower scores for athletes competing closely against their compatriots (Ansorge & Scheer, 1988; Braeunig, 2024; Heiniger & Mercier, 2021). A third form, "indirect" (Sandberg, 2018, p. 2146) national bias, occurs when judges adjust their scores to compensate for or align with the perceived or anticipated national bias of their colleagues (Sandberg, 2018). While patriotic bias has been consistently documented across multiple sports, including Figure Skating (Sala et al., 2007; Zitzewitz, 2006, 2014), Ski Jumping (Bouwens et al., 2022; Lyngstad et al., 2020), Gymnastics (Heiniger & Mercier, 2019, 2021), Diving (Emerson et al., 2009), Dressage (Sandberg, 2018), and Muay Thai (Myers et al., 2006), competitor bias and indirect national bias have received considerably less scholarly attention. To date, only three studies have reported evidence for competitor bias (Ansorge & Scheer, 1988; Braeunig, 2024; Heiniger & Mercier, 2021) and indirect national bias respectively (Sandberg, 2018; Wolframm, 2023; Zitzewitz, 2006).

Importantly, prior research suggests that the magnitude of patriotic bias varies substantially between individual judges. While some judges demonstrate strong favoritism toward compatriots, others remain neutral or even penalize athletes from their own country (Heiniger & Mercier, 2019; Sandberg, 2018). Moreover, national bias has been found to intensify in later competition stages, where higher stakes such as medal contention are at play (Heiniger & Mercier, 2019, 2021; Zitzewitz, 2006). However, this relationship remains contested, with several studies finding no significant association between competition stage and the magnitude of bias (Campbell & Galbraith, 1996; Krumer et al., 2022; Morgan & Rotthoff, 2014; Sala et al., 2007; Sandberg, 2018; Scholten et al., 2020). Finally, it has been hypothesized that national bias may be more pronounced in highly subjective performance aspects, where evaluative discretion is greater (Fenwick & Chatterjee, 1981; Lee, 2008). Empirical findings, however, remain inconsistent (Lee, 2008; Sandberg, 2018; Yang, 2006; Zitzewitz, 2006, 2014) and are often based on nominal comparisons of bias estimates rather than rigorous statistical testing (Fang & Ho, 2024; Fenwick & Chatterjee, 1981; Lee, 2008; Zitzewitz, 2006).

Taken together, aesthetic sports face persistent challenges associated with both actual and perceived judging errors and biases that threaten the integrity of competition outcomes, the credibility of judges and judging systems, and the legitimacy of their organizational governance. However, the distinctive competition format and comparative scoring system of Breaking may be particularly susceptible to errors and biases. The sport's Olympic debut therefore offers a valuable opportunity to empirically examine both interrater reliability and national bias within Breaking's unique competitive and evaluative framework.

Building on insights from prior research, the present study proposes and subsequently tests the following hypotheses:

- **H1**: Olympic Breaking judges exhibit national bias by awarding inflated scores to athletes from their own country.
- **H2**: In Olympic Breaking, national bias is more pronounced when the nationalities of both athletes involved in a head-to-head battle are represented on the judging panel (subsequently referred to as *Both-Athletes-Represented* (*BAR*)).
- **H3**: The magnitude of national bias increases in later stages of Olympic Breaking events, where stakes are higher than in early stages.
- **H4**: The magnitude of national bias in Olympic Breaking varies across the five performance aspects.

In addition, this study assesses interrater reliability in Olympic Breaking, contributing to the broader discourse on judging reliability and objectivity in aesthetic sports, and providing the first empirical assessment of interrater reliability within a sport's scoring system based on direct performance comparisons rather than consecutive and independent performance evaluations.

## 2 Methods

### 2.1 Scoring Data

The scoring data for the 2024 Olympic Breaking events were sourced from the official *Book of Results* (Paris Organising Committee for the 2024 Olympic and Paralympic Games, 2024). This dataset includes all judges' scores from both the B-Boys and B-Girls events, as well as the nationalities of the athletes and judges. In total, it contains 7,776 scores—3,888 from each event. All scores were issued by the same panel of nine judges who evaluated 32 athletes (16 B-Boys and 16 B-Girls). A descriptive summary of the scoring data is provided in Table 1.

Table 1:
Summary of scoring data from the 2024 Olympic Breaking competition.

|                              | Overall | B-Boys | B-Girls |
|------------------------------|---------|--------|---------|
| Judges                       | 9       | 9      | 9       |
| Athletes                     | 32      | 16     | 16      |
| Battles                      | 64      | 32     | 32      |
| **Scores**                   | **7,776** | **3,888** | **3,888** |
| thereof for Compatriots      | 996     | 480    | 516     |
| thereof for Compatriots in BAR | 588   | 324    | 264     |
| thereof for non-Compatriots  | 6,780   | 3,408  | 3,372   |

In Olympic Breaking, judges evaluated between two and three rounds per battle, depending on the competition stage. Each battle featured two athletes, who alternated performances, with one initiating the round and the other responding. Judges assessed each round across five performance aspects—*Technique*, *Vocabulary*, *Originality*, *Execution*, and *Musicality*—and assigned a single comparative score for each aspect. These scores represent the perceived quality difference between the two athletes on a continuous scale from 0% to 20%, in 0.2% increments. Judges indicated which athlete they deemed superior using color codes (red or blue), corresponding to the respective athlete (see Figure 1). The five performance aspect scores were subsequently aggregated into an overall performance difference score for each judge, which directly determined the judge's vote for the round winner (red or blue). The round winner was decided by a majority of votes across the nine judges. The athlete who won the most rounds was declared the battle winner. In the case of a tie in rounds, the cumulative number of judge votes served as a tiebreaker (Paris Organising Committee for the 2024 Olympic and Paralympic Games, 2024; WDSF, 2025).

Both events followed a structured progression format, beginning with a Round Robin stage in which four athletes competed against each other within four separate groups. The top two athletes from each group advanced to further stages, comprising the Quarterfinals, Semifinals, Bronze Medal Battle, and Gold Medal Battle. These stages followed a knockout format, with athletes progressing based on battle victories (Paris Organising Committee for the 2024 Olympic and Paralympic Games, 2024).

For subsequent analyses, the categorical color indicators (red and blue) were transformed into numerical values by assigning positive scores to red athlete superiority votes and negative scores to blue athlete superiority votes. This transformation allowed continuous data analysis independent of nominal color labels. A score of 0.0% represents no perceived performance difference between athletes, denoted as "–" in the official dataset.

| | | Battle ███ | | | |
|---|---|---|---|---|---|
| Red side ███ ███ | 🟥 | 1-1 12-6 (8-1, 4-5) | 🟦 | Blue side ███ | |

**Round 1** 🟥 Red  1-0 (8-1)

| Judge | Overall | Technique | Vocabulary | Originality | Execution | Musicality |
|---|---|---|---|---|---|---|
| ███ | Red +5.8% | Red +1.6% | Red +2.2% | Blue +1.4% | Red +1.6% | Red +1.8% |
| ███ | Red +13.0% | Red +5.4% | Red +3.4% | Blue +4.4% | Red +5.0% | Red +3.6% |
| ███ | Blue +4.6% | Blue +1.2% | Blue +1.0% | Blue +1.8% | Red +2.0% | Blue +2.6% |
| ███ | Red +10.0% | Red +2.8% | Red +3.2% | Blue +1.6% | Red +3.8% | Red +1.8% |
| ███ | Red +2.2% | Red +0.8% | Red +0.6% | Blue +0.8% | Red +0.6% | Red +1.0% |
| ███ | Red +5.0% | Red +1.6% | Red +2.2% | Blue +1.2% | Red +1.2% | Red +1.2% |
| ███ | Red +1.8% | Red +2.6% | Blue +2.6% | Red +2.2% | Blue +2.2% | Red +1.8% |
| ███ | Red +6.2% | Red +7.8% | Blue +4.0% | Red +3.0% | Blue +3.8% | Red +3.2% |
| ███ | Red +9.6% | Red +2.0% | Red +2.8% | Red +0.4% | Red +1.8% | Red +2.6% |

**Round 2** 🟦 Blue  1-1 (4-5)

| Judge | Overall | Technique | Vocabulary | Originality | Execution | Musicality |
|---|---|---|---|---|---|---|
| ███ | Blue +0.6% | Red +1.8% | Blue +1.2% | Blue +1.6% | Blue +0.8% | Red +1.2% |
| ███ | Blue +4.0% | Red +2.2% | Blue +2.2% | Blue +4.6% | Blue +1.2% | Red +1.8% |
| ███ | Red +1.6% | Red +2.4% | Blue +0.6% | Blue +0.4% | Red +1.4% | Blue +1.2% |
| ███ | Blue +6.6% | Blue +1.6% | Blue +1.8% | Blue +3.2% | Blue +1.4% | Red +1.4% |
| ███ | Red +1.8% | Red +1.0% | Red +1.0% | Blue +1.2% | - | Red +1.0% |
| ███ | Blue +7.6% | Blue +1.6% | Blue +1.2% | Blue +1.8% | Blue +1.4% | Blue +1.6% |
| ███ | Red +5.0% | Red +1.4% | Red +1.2% | Red +0.8% | Red +1.2% | Red +0.4% |
| ███ | Red +3.4% | Red +3.6% | Red +3.0% | Blue +2.8% | Blue +3.4% | Red +3.0% |
| ███ | Blue +4.0% | Blue +1.0% | Blue +2.4% | Blue +1.0% | Blue +0.8% | Red +1.2% |

*Fig. 1* Excerpt (screenshot) from the official scoring report for the 2024 Olympic Breaking events (Paris Organising Committee for the 2024 Olympic and Paralympic Games, 2024), illustrating the scoring results of a single battle comprising two rounds. The names of judges and athletes, as well as the battle number, were redacted following the capture of the excerpt.

## 2.2 Interrater Reliability Analyses

Interrater reliability among Breaking judges was assessed using the Intraclass Correlation Coefficient (ICC), a recognized and appropriate measure for evaluating reliability among multiple judges in aesthetic sports (Klein et al., 2014; Wirtz & Caspar, 2002). Although the ICC has been widely used to assess the interrater reliability of sports judges, prior studies often failed to specify the ICC variant employed or applied inappropriate models of the ICC for the given scoring data or research objectives.

In Olympic Breaking, the same nine judges evaluate all athlete performances throughout the competition. Consequently, the two-way random model of the ICC is employed to assess interrater reliability. It assumes that a fixed set of raters conducts all evaluations and permits generalization to a larger population of raters (Shrout & Fleiss, 1979; Wirtz & Caspar, 2002).

Because judges in Breaking assess relative performance differences between two athletes rather than individual performances, no meaningful systematic mean score differences are assumed among the individual judges. Furthermore, unlike in combat sports, where color-coded uniforms have been shown to influence scoring (Hill & Barton, 2005; Sorokowski et al., 2014), such effects are unlikely in Breaking, as athletes do not wear designated red or blue attire. Under these conditions, where no systematic scoring differences among judges are expected, Wirtz and Caspar (2002) recommend the unadjusted ICC variant. It is less restrictive in its interpretation and provides estimates comparable to the adjusted ICC variant. Accordingly, the single-measures ICC ($ICC_s$) based on the unadjusted two-way random model was calculated as follows to assess interrater reliability in Olympic Breaking (Wirtz & Caspar, 2002):

$$ICC_s(A, 1) = \frac{\sigma_p^2}{\sigma_p^2 + \sigma_j^2 + \sigma_{pj}^2 + \sigma_\varepsilon^2} \tag{0}$$

Here, $\sigma_p^2$ represents variance of the judges' scores attributed to the differences in athlete performances ($p$), while the denominator comprises total score variance, including variance attributed to the differences in athlete performances ($\sigma_p^2$), systematic scoring differences among judges ($\sigma_j^2$), judge-performance interaction variance ($\sigma_{pj}^2$), and residual error variance ($\sigma_\varepsilon^2$). The resulting $ICC_s(A, 1)$ value reflects the proportion of scoring variance solely attributable to differences in performances, ranging from 0 to 1. A value of 1 thereby indicates perfect agreement among judges, signifying that scoring variance is entirely explained by performance differences (McGraw & Wong, 1996; Shrout & Fleiss, 1979; Wirtz & Caspar, 2002).

In Olympic Breaking, the progression of athletes through the events is determined solely by the outcomes of individual battles. Comparisons between battles are not conducted, and judges are required only to identify the winner of each battle round. As a result, $ICC_s(A, 1)$ estimates calculated using scores from multiple battles (henceforth called *Overall ICC* measures) are likely to overestimate the interrater reliability in Breaking. This overestimation stems from between-battle scoring variance inflating $\sigma_p^2$ and thus ICC values, despite having no influence on competition outcomes.

Nevertheless, overall ICCs remain useful when assessing subsets of the data (e.g., scoring of individual performance aspects or specific rounds) where meaningful within-battle ICCs cannot be estimated due to limited data points (2–3 scores per judge and battle for individual performance aspects). Consequently, assessments of interrater reliability for separate performance aspects or battle rounds are conducted independent of individual battles, which is possible because the panel composition remained stable across the Olympic Breaking events. The interpretation of these overall ICC measures, however, requires caution, as they likely overstate the actual interrater reliability in this context.

## 2.3 Estimation of National Bias in Breaking

### 2.3.1 Estimating Overall National Bias

Given the comparative nature of scoring in Olympic Breaking, it is not possible to differentiate between patriotic bias and competitor bias. To evaluate the presence and magnitude of an overall national bias (Hypothesis 1)—encompassing both forms—a linear mixed-effects regression modelling approach was employed, consistent with established methods in prior research on national bias in sports judging (Krumer et al., 2022; Litman & Stratmann, 2018; Sandberg, 2018; Zitzewitz, 2006). The model estimating overall national bias utilized only the judges' overall scores (aggregated performance aspect scores) and is specified as follows:

$$S_{jpa_{(r,b)}} = \theta_{pa_{(r,b)}} + \beta_{NB}\, \phi(Nat_{a_r} = Nat_j \oplus Nat_{a_b} = Nat_j) \quad (1)$$

Here, $S_{jpa_{(r,b)}}$ denotes the score assigned by judge $j$ for a specific performance comparison $p$ involving the athlete pair $a_{(r,b)}$. Each performance comparison $p$ corresponds to a distinct round within a specific battle in either the B-Boys or B-Girls event, thereby identifying all unique performance comparisons evaluated during the 2024 Olympic Breaking events. The performance fixed effect $\theta_{pa_{(r,b)}}$ approximates the true performance difference between the two athletes $a_{(r,b)}$ for each specific performance comparison $p$.

The dummy variable $\phi(Nat_{a_r} = Nat_j \oplus Nat_{a_b} = Nat_j)$ captures instances in which judge $j$ shared the same nationality with either the red or blue athlete ($a_r$ XOR $a_b$). Reflecting the numerical coding of athlete identifiers (red and blue), this variable takes a value of 1 when the judge shares the red athlete's nationality, −1 when sharing the blue athlete's nationality, and 0 when sharing both or neither athlete's nationality. The coefficient $\beta_{NB}$ thus represents the estimated magnitude of national bias exhibited by the judges.

### 2.3.2 Estimating Indirect National Bias

Building upon Model (1), an extended Model (2) incorporates an additional dummy variable $\phi(BAR)$ to capture indirect national bias. Prior research suggests that judges may adapt their scoring behavior based on perceived or anticipated bias by other judges (Hypothesis 2). To examine this potential behavioral adjustment, the additional dummy variable identifies scoring cases in which both athletes were represented by different judges on the panel (BAR):

$$S_{jpa_{(r,b)}} = \theta_{pa_{(r,b)}} + \beta_{NB}\, \phi(Nat_{a_r} = Nat_j \oplus Nat_{a_b} = Nat_j) + \beta_{BAR}\, \phi(BAR) \quad (2)$$

In this model, $\beta_{BAR}$ estimates the incremental effect of BAR conditions on overall national bias, while $\beta_{NB}$ remains as a baseline estimate of (direct) national bias.

## 2.3.3 Estimating the Effect of Competition Progression on National Bias

To assess whether national bias changed as the events advanced (Hypothesis 3), a variable representing the competition stages (*Stage*) of Breaking events is incorporated into Model (1). This variable represents the stage in which the score $S_{jpa_{(r,b)}}$ was assigned, with increasing integer values reflecting event progression (Round Robin Stage = 0, Quarterfinals = 1, Semifinals = 2, Bronze/Gold Battles = 3):

$$S_{jpa_{(r,b)}} = \theta_{pa_{(r,b)}} + \beta_{NB}\, \phi(Nat_{a_r} = Nat_j \oplus Nat_{a_b} = Nat_j) \\ + \beta_S\, Stage{:}\phi(Nat_{a_r} = Nat_j \oplus Nat_{a_b} = Nat_j) \qquad (3)$$

Here, $\beta_{NB}$ captures the baseline national bias, while $\beta_S$ reflects how the magnitude of national bias evolves as competition intensity and stakes rise across the above-defined successive stages of the Olympic Breaking events.

## 2.3.4 Estimating National Bias across different Performance Aspects

To examine whether the extent of national bias varied across the five performance aspects (Hypothesis 4), Model (1) was again extended and applied to a dataset comprising the judges' individual performance aspect-specific scores rather than their overall scores. Separate and mutually exclusive identification variables, $\phi(Aspect)$, were created for each performance aspect, coded as 1 when the score $S_{jp_A a_{(r,b)}}$ corresponded to that specific performance aspect. Each aspect-specific performance comparison ($p_A$) thus represents a unique performance comparison in which a particular performance aspect ($A$) was evaluated in a distinct round of a specific battle within either the B-Boys or B-Girls competition.

Within this extended model, the performance aspect identification variables each interact with a national bias dummy variable. The model is specified as follows:

$$S_{jp_A a_{(r,b)}} = \theta_{p_A a_{(r,b)}} + \beta_A\, \phi(Aspect){:}\phi(Nat_{a_r} = Nat_j \oplus Nat_{a_b} = Nat_j) \qquad (4.1)$$

The coefficient $\beta_A$ is aspect-specific and quantifies the magnitude of national bias estimated for each distinct performance aspect.

To further compare national bias across aspects, the scoring dataset containing the individual performance aspect scores was divided into subsets containing all scores related to two performance aspects at a time, enabling pairwise comparisons. An interaction term between $\phi(Nat_{a_r} = Nat_j \oplus Nat_{a_b} = Nat_j)$ and one specific performance aspect indicator $\phi(Aspect)$ was introduced into Model (1). This modification allowed for the pairwise comparison of national bias estimates across the two performance aspects contained in the respective scoring data subsets:

$$S_{jp_A a_{(r,b)}} = \theta_{p_A a_{(r,b)}} + \beta_\Delta \, \phi(Nat_{a_r} = Nat_j \oplus Nat_{a_b} = Nat_j) \\ + \beta_A \, \phi(Aspect){:}\phi(Nat_{a_r} = Nat_j \oplus Nat_{a_b} = Nat_j) \quad (4.2)$$

Here, $\beta_\Delta$ represents the difference in national bias between the two performance aspects under comparison, whereas $\beta_A$ quantifies the magnitude of national bias for the specific performance aspect indicated by the interaction. Pairwise comparisons were conducted across all five performance aspects, with an adjusted significance level reported alongside the corresponding results.

### 2.3.5 Exploring National Bias in B-Boys vs. B-Girls Events

To explore potential differences in national bias between the B-Boys and B-Girls events, two mutually exclusive event indicators ($\phi$(*B-Boys*), $\phi$(*B-Girls*)) were incorporated into Model (1), each interacting with a national bias dummy variable:

$$S_{jpa_{(r,b)}} = \theta_{pa_{(r,b)}} + \beta_{BB} \, \phi(B\text{-}Boys){:}\phi(Nat_{a_r} = Nat_j \oplus Nat_{a_b} = Nat_j) \\ + \beta_{BG} \, \phi(B\text{-}Girls){:}\phi(Nat_{a_r} = Nat_j \oplus Nat_{a_b} = Nat_j) \quad (5.1)$$

This model, based on the judges' overall scores, provides event-specific estimates of overall national bias for both the B-Boys ($\beta_{BB}$) and the B-Girls ($\beta_{BG}$) events.

By removing the *B-Boys* event indicator from the first interaction term in Model (5.1), the coefficient of the sole national bias dummy variable ($\beta_\Delta$) captures the difference in national bias between the two event categories, while $\beta_{BG}$ represents the national bias specific to the B-Girls event:

$$S_{jpa_{(r,b)}} = \theta_{pa_{(r,b)}} + \beta_\Delta \, \phi(Nat_{a_r} = Nat_j \oplus Nat_{a_b} = Nat_j) \\ + \beta_{BG} \, B\text{-}Girls{:}\phi(Nat_{a_r} = Nat_j \oplus Nat_{a_b} = Nat_j) \quad (5.2)$$

### 2.3.6 Exploring judge-specific National Bias

Finally, to explore whether overall national bias is uniformly exhibited across judges or primarily driven by specific individuals, judge-specific dummy variables $\phi(j)$ were introduced into Model (1) and interacted with the national bias dummy variable:

$$S_{jpa_{(r,b)}} = \theta_{pa_{(r,b)}} + \beta_j \phi(j){:}\phi(Nat_{a_r} = Nat_j \oplus Nat_{a_b} = Nat_j) \quad (6)$$

The resulting coefficient $\beta_j$ provides judge-specific estimates of national bias, thereby enabling an assessment of whether certain judges disproportionately contributed to overall national bias.

All statistical analyses were conducted using R Statistical Software (v4.3.2; R Core Team, 2023). Dataset management and preprocessing were performed using the *dplyr* package (v1.1.4; Wickham et al., 2023). Intraclass correlation coefficients were computed using the *irr* package (v0.84.1; Gamer & Lemon, 2019), and linear mixed-effects regression models were estimated with the *fixest* package (v0.12.1; Bergé, 2018). P-values associated with Hypotheses H1–H3 were adjusted for one-sided testing following model estimation.

# 3 Results

## 3.1 Interrater Reliability in Olympic Breaking

Table 2 presents the results of the interrater reliability analysis, examining judge reliability across individual events, performance aspects, rounds within battles, and competition stages. Negative ICC values are interpreted as indicating a reliability of 0.0, following Wirtz and Caspar (2002). Overall, the findings indicate low to moderate levels of interrater reliability according to established interpretive standards (Wirtz & Caspar, 2002). Thus, the performance assessments in Breaking are notably less reliable than in other judged sports, where reliability levels typically range from moderate to very high (Braeunig, 2025).

Table 2:
Summary of interrater reliability results.

|  |  | $ICC_s(A, 1)$ | | | | | |
|---|---|---|---|---|---|---|---|
| Subset | Event | Within individual Battles | | | | | Overall ICC |
|  |  | n Battles | Range | P25 | Median | P75 | Mean ± SD* |  |
| All Aspects | B-Boys & B-Girls | 64 | -.027 – .611 | .060 | .156 | .230 | .174 ± .150 | .336 |
| All Aspects | B-Boys | 32 | .003 – .611 | .083 | .131 | .230 | .178 ± .153 | .320 |
| All Aspects | B-Girls | 32 | -.027 – .552 | .042 | .168 | .213 | .170 ± .146 | .340 |
| Overall Scores | B-Boys & B-Girls | 64 |  |  |  |  |  | .425 |
| Technique | B-Boys & B-Girls | 64 |  |  |  |  |  | .348 |
| Vocabulary | B-Boys & B-Girls | 64 |  |  |  |  |  | .320 |
| Originality | B-Boys & B-Girls | 64 |  |  |  |  |  | .326 |
| Execution | B-Boys & B-Girls | 64 |  |  |  |  |  | .371 |
| Musicality | B-Boys & B-Girls | 64 |  |  |  |  |  | .318 |
| Battle Round 1 | B-Boys & B-Girls | 64 |  |  |  |  |  | .318 |
| Battle Round 2 | B-Boys & B-Girls | 64 |  |  |  |  |  | .363 |
| Battle Round 3 | B-Boys & B-Girls | 16 |  |  |  |  |  | .284 |
| Round Robin | B-Boys & B-Girls | 48 | -.027 – .611 | .050 | .138 | .200 | .164 ± .150 | .354 |
| Quarterfinals | B-Boys & B-Girls | 8 | .003 – .442 | .083 | .210 | .221 | .217 ± .158 | .296 |
| Semifinals | B-Boys & B-Girls | 4 | .081 – .437 | .084 | .143 | .259 | .206 ± .156 | .230 |
| Finals | B-Boys & B-Girls | 4 | .024 – .319 | .156 | .214 | .250 | .195 ± .110 | .247 |

*Mean ± SD values were calculated using a nonlinear Fisher *z*-transformation and subsequently back-transformed to the original scale, following Wirtz and Caspar (2002).

## 3.2 National Bias in Olympic Breaking

Table 3 summarizes the results of national bias analyses across Models (1)–(6).

Model (1) reveals a statistically significant overall tendency for judges to favor athletes of their own nationality. On average, compatriot athletes received 2.47% higher overall scores than non-compatriot athletes ($\beta_{NB}$ = 2.474; t(1,295) = 6.155; p < 0.001).

Model (2) incorporated a control for cases in which both competing athletes' nationalities were represented by judges on the panel (BAR). The inclusion of this control did not indicate a statistically significant change in the level of national bias ($\beta_{BAR}$ = 0.379; t(1,294) = 0.467; p = 0.641). When both athletes were represented by judges, the bias increased only marginally to 2.62%.

Model (3) examined the relationship between competition progression and national bias. The results indicate no significant increase in national bias as the events progressed ($\beta_S$ = 0.027; t(1,294) = 0.079; p = 0.937), indicating that national bias remained stable across stages. This suggests that judges did not strategically increase or suppress bias during later and more consequential stages of the events, such as medal-deciding battles.

Model (4.1) estimated national bias across the five performance aspects. National bias was most pronounced in the scoring of the *Technique* and *Musicality* aspects, with additional significant findings for the *Execution* aspect (Figure 2). Pairwise comparisons conducted via Model (4.2) revealed significant differences in national bias between the performance aspects *Musicality* and *Vocabulary* ($\beta_\Delta$ = 0.668; t(2,590) = 3.175; p = 0.002), and between *Musicality* and *Originality* ($\beta_\Delta$ = 0.630; t(2,590) = 3.153; p = 0.002). A summary of results for all pairwise comparisons is reported in Table 4.

Model (5.1) estimated national bias separately for the B-Boys and B-Girls events. Statistically significant national bias was detected in both events ($\beta_{BB}$ = 2.155; t(1,294) = 3.389; p < 0.001 / $\beta_{BG}$ = 2.776; t(1,294) = 5.605; p < 0.001); however, the difference in national bias between the events was not statistically significant according to Model (5.2) ($\beta_\Delta$ = 0.621; t(1,294) = 0.770; p = 0.442).

Model (6) explored judge-specific national bias. The results show that the estimated overall national bias was not uniformly exhibited across judges but was instead driven by a small subset of judges. Two judges—Judge 6 and Judge 8—displayed significant favoritism toward compatriot athletes, awarding average advantages of 4.35% and 5.47%, respectively. Three additional judges exhibited positive but statistically insignificant bias. In contrast, one judge (Judge 4) demonstrated a negative bias of –4.06%, indicating a tendency to disadvantage compatriots relative to non-compatriots.

Table 3:
Summary of results from Models (1)–(6).

|  | Model 1 | Model 2 | Model 3 | Model 4.1* | Model 5.1 | Model 6 |
|---|---|---|---|---|---|---|
| **National Bias** | **2.474*** (6.155)** | **2.239*** (3.612)** | **2.457*** (5.177)** | | | |
| National Bias in BAR | | 0.379 (0.467) | | | | |
| Stage:National Bias | | | 0.027 (0.079) | | | |
| **Technique** | | | | **0.715*** (4.181)** | | |
| Vocabulary | | | | 0.242 (1.696) | | |
| **Originality** | | | | **0.281* (2.221)** | | |
| **Execution** | | | | **0.326* (2.153)** | | |
| **Musicality** | | | | **0.911*** (5.913)** | | |
| **B-Boys Event** | | | | | **2.155*** (3.389)** | |
| **B-Girls Event** | | | | | **2.776*** (5.605)** | |
| Judge 1 | | | | | | NA† |
| **Judge 2** | | | | | | **1.740*** (3.561)** |
| Judge 3 | | | | | | NA† |
| Judge 4 | | | | | | -4.801 (-1.910) |
| **Judge 5** | | | | | | **2.034* (2.400)** |
| **Judge 6** | | | | | | **4.746*** (4.313)** |
| Judge 7 | | | | | | 0.984 (1.416) |
| **Judge 8** | | | | | | **5.849*** (4.510)** |
| Judge 9 | | | | | | NA† |
| Adjusted $R^2$ | 0.433 | 0.432 | 0.432 | 0.342 | 0.432 | 0.437 |
| Within $R^2$ | 0.016 | 0.016 | 0.016 | 0.010 | 0.016 | 0.028 |

*Model (4.1) was estimated exclusively using scores for individual performance aspects rather than overall scores.

†No scores were provided for compatriot athletes.

Significant effects are displayed in **bold**, with *t*-statistics shown in parentheses: *p < 0.05, ***p < 0.001

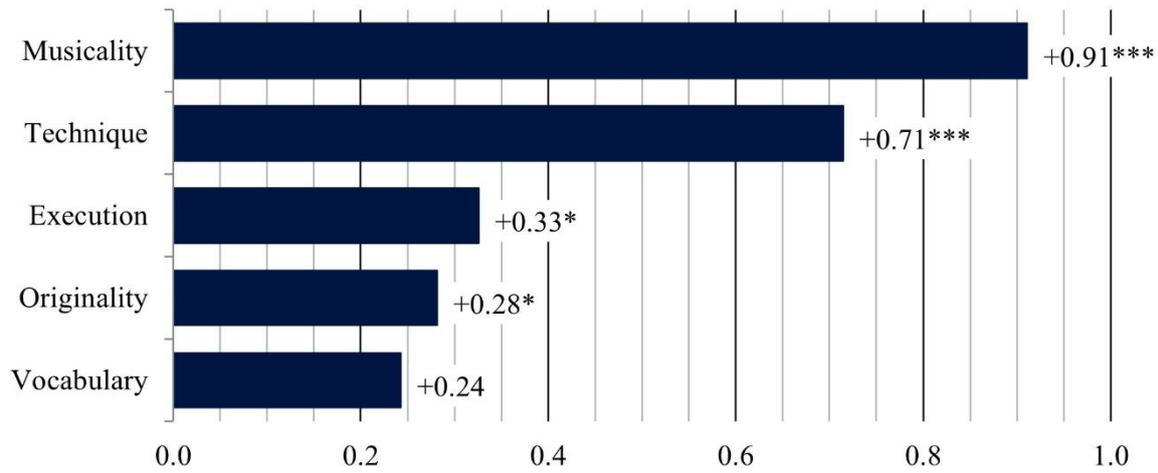

**Fig. 2** Estimated magnitude of national bias across performance aspects. *p < 0.05, ***p < 0.001

Table 4:
Summary of pairwise comparisons of national bias across performance aspects based on Model (4.2)

| Pairwise Comparison | $\beta_\Delta$ | t | p* |
| --- | --- | --- | --- |
| Vocabulary - Technique | -0.4718 | -2.1142 | 0.0354 |
| Originality - Technique | -0.4331 | -2.0334 | 0.0429 |
| Execution - Technique | -0.3885 | -1.6995 | 0.0903 |
| Musicality - Technique | 0.1965 | 0.8530 | 0.3944 |
| Originality - Vocabulary | 0.0387 | 0.2023 | 0.8398 |
| Execution - Vocabulary | 0.0833 | 0.3994 | 0.6899 |
| **Musicality - Vocabulary** | **0.6683** | **3.1749** | **0.0017** |
| Execution - Originality | 0.0446 | 0.2256 | 0.8217 |
| **Musicality - Originality** | **0.6296** | **3.1528** | **0.0018** |
| Execution - Musicality | 0.5850 | 2.7050 | 0.0072 |

*p-values are unadjusted for multiple comparisons; according to the conservative Bonferroni correction, a significance threshold of p < 0.005 applies.

## 3.3 Effects of National Bias on the Outcomes of Olympic Breaking Events

Lastly, an exploratory analysis examined the potential influence of national bias on the outcomes of the 2024 Olympic Breaking events. Using the judge-specific bias estimates derived from Model (6), instances were identified in which a judge's vote could have been altered in favor of a compatriot athlete due to national bias.

Out of 166 total judge votes involving compatriot athletes, 20 votes (12.0%) were identified as potentially altered by national bias based on the individual judges overall scores. Among these 20 votes, two instances were found in which the presumed bias-driven vote alteration directly influenced the round outcome within a battle. In both cases, the overall panel vote was estimated to shift from 4–5 to 5–4 in favor of the compatriot athlete.

In one of these instances, the presumed alteration also affected the overall battle result and consequently the advancement of athletes within the round-robin stage, allowing the respective compatriot athlete to progress to the quarterfinals stage in place of another competitor. In the second instance, national bias may have influenced one round of a gold medal battle; however, this did not alter the final result, as the respective compatriot athlete had already secured a 2–0 round lead prior to the affected outcomes of Round 3.

## 4 Discussion

This study examined interrater reliability and potential national bias among judges in Olympic Breaking. Building on prior research in aesthetic sports, it was hypothesized that Breaking judges would demonstrate a tendency to favor compatriot athletes (Hypothesis 1). Additional hypotheses addressed the possibility of indirect national bias when both athletes' nationalities were represented on the judging panel (Hypothesis 2), a potential increase in national bias as the B-Boy and B-Girl events progressed (Hypothesis 3), and potential differences in the magnitude of national bias across various performance aspects (Hypothesis 4).

The results revealed low to moderate interrater reliability among judges in Olympic Breaking, with ICC values ranging from –0.03 to 0.611. Reliability levels were consistent across performance aspects, battle rounds, and competition stages, suggesting that limited agreement among judges is a general feature of performance evaluation in Breaking. Contrary to prior findings, judges did not show greater inconsistency in the assessment of presumably more subjective performance aspects, implying that the observed low reliability reflects systemic challenges within the judging process. Compared with multiple other aesthetic sports (Aleksić-Veljković et al., 2016; Bučar et al., 2011; Bučar et al., 2013; Bučar Pajek et al., 2011; Klein et al., 2014; Leandro et al., 2017; Leskošek et al., 2010; Leskošek et al., 2012; Leskošek et al., 2018), the level of reliability in Olympic Breaking was notably lower, and it also fell slightly below that reported in dance-related sports (Pavleski & Bozilova, 2020; Premelč et al., 2019; Sato, 2022; Sato & Hopper, 2021). While prior limitations in reliability estimation limit direct comparisons, these findings nevertheless suggest a need for improvement in Breaking's judging.

A possible explanation for the low reliability lies in Breaking's unique competition and scoring format, in which two performances are evaluated directly against each other. This format may impose high cognitive demands on judges, who must memorize a detailed account of one athlete's performance within five separate performance aspects, until they provide one comparative score per performance aspect after the conclusion of the second athlete's performance. Thus, a potential refinement could involve scoring each individual performance independently, followed by a comparative evaluation based on the individual scores. Such a modification may reduce cognitive load and enhance both transparency and scoring reliability (Bailie, 1965; Landers, 1970; Ste-Marie, 2000; Wolframm, 2023). Moreover, the findings emphasize the importance of clear evaluative criteria and a shared understanding of performance quality within the Breaking community to enhance judging quality. As with other aesthetic sports, the ongoing negotiation of what constitutes a high-quality performance is an essential part of the sport's maturation (Stern, 2010). Evidence of this process emerged during the 2024 Paris Olympic Games, when Australian breaker Rachael Gunn ("RAYGUN") received the lowest possible jury votes. In an interview with *The Guardian* (2024), Gunn described her performance as a form of art, noting that a creative approach might resonate with the jury—or it might not. The ensuing public debate about the artistic versus athletic interpretation of Breaking performances illustrates how evolving community discourse may gradually strengthen the common conceptual basis for judging and, in turn, may improve scoring reliability as evaluation criteria become more clearly defined.

The analysis of national bias revealed that judges awarded, on average, 2.36% higher overall scores to athletes of their own nationality, representing a substantial and statistically significant national bias and thus supporting Hypothesis 1. However, the presence of a judge representing the opposing athlete on the panel (BAR condition) did not significantly affect the magnitude of this bias, nor did bias intensify as the events advanced. Consequently, Hypotheses 2 and 3 are not supported by the results. This suggests that Breaking judges do not manipulate scores strategically, contrasting with findings from figure skating and ski jumping, where strategic score manipulation to counteract or affirm each other's national bias has been documented (Sandberg, 2018; Zitzewitz, 2006). Consistent with Hypothesis 4, significant differences in national bias emerged across performance aspects: it was most pronounced in the evaluation of *Technique* and *Musicality*, and to a lesser extent in *Execution*. These findings may align with prior research indicating that more interpretive or subjective performance aspects are more vulnerable to evaluative bias (Fang & Ho, 2024; Fenwick & Chatterjee, 1981; Lee, 2008; Sandberg, 2018; Yang, 2006; Zitzewitz, 2006, 2014). However, the interrater reliability analysis did not reveal notable differences in reliability across these aspects, suggesting no substantial variation in subjectivity. Thus, national bias cannot be clearly attributed to differences in assessment subjectivity between performance aspects.

Exploratory analyses further revealed no significant differences in national bias between the B-Boys and B-Girls events, although a marginal difference of 0.62% in national bias magnitude persisted in overall scores. Judge-specific analyses indicated that national bias was not uniformly exhibited across the judges but primarily driven by two individuals who exhibited substantial favoritism, awarding compatriot athletes up to 5.5% higher overall scores compared to non-compatriots. In contrast to sports like figure skating and ski jumping, where national bias appears more widespread among judges (Braeunig, 2024; Heiniger & Mercier, 2019; Sandberg, 2018), these findings suggest that bias in Breaking may stem from a narrow subset of judges rather than representing a systemic panel-wide phenomenon. Interestingly, one judge exhibited a notable, but statistically non-significant tendency to penalize compatriot athletes, highlighting potential variability in the judges' abilities to navigate scoring margins strategically (Braeunig, 2024).

Despite the considerable presence of national bias in Olympic Breaking, its practical influence on competition outcomes was limited, potentially influencing competition standings in only a single instance. Exploratory analyses identified only two battle rounds (1.2% of judge votes involving compatriots) where the estimated national bias may have affected results, with just one case potentially altering the final competition standings. This minimal effect can be attributed to Breaking's scoring format, which aggregates individual judges' scores into binary votes. Such a system inherently constrains the impact of any single judge's bias, as distortions can only influence outcomes in narrowly contested vote distributions (e.g., 4–5 situations) or when overall vote counts are used as tiebreakers in the initial group phase of the competition. Unlike in many aesthetic sports where aggregated score averages directly determine results, Breaking's scoring system offers a built-in safeguard against disproportionate effects of individual bias. Nonetheless, this design may introduce alternative vulnerabilities: the limitation of an individual judge's ability to alter competition outcomes in Breaking could, in cases of intentional bias or partiality, incentivize collusion through vote trading as a means of exerting a more decisive influence on results.

Several limitations of this study must be considered. Although the dataset included over 6,000 individual scores, the analysis was limited to a single Olympic competition encompassing two events and nine judges. This narrow scope constrains the generalizability of the findings, though it represents a complete survey of Olympic-level judging in Breaking to date. Moreover, as all judges were certified to officiate at the Olympic level, the results may not reflect judging behavior at lower competition levels or among less experienced judges. Consequently, the results may not be applicable to less experienced judges or those officiating in less prestigious competitions. Finally, the estimation of national bias effects on competition outcomes based on judge-specific national bias estimates is inherently constrained. Because these estimates represent average scoring tendencies derived from numerous scoring instances, they may not fully capture the variability of national bias at the level of individual scoring instances, where its magnitude and influence may fluctuate.

Future research should therefore extend the scope to include diverse Breaking competitions and events, featuring a broader range of expertise among judges. Additionally, analyses that track how judging reliability and bias evolve as Breaking further develops may provide valuable insights into how scoring systems and a common conceptual basis influence scoring reliability and bias.

# 5 Conclusion

This study revealed low to moderate interrater reliability among judges in the 2024 Olympic Breaking events, raising concerns about the employed scoring system, the judges' ability to consistently discern quality differences between Breaking performances, and potential inconsistencies in the judges' interpretation and application of the performance quality assessment criteria. As a result, competition outcomes may have been considerably influenced by random scoring variation or systematic discrepancies in how judges applied scoring criteria and guidelines.

Additionally, two judges demonstrated statistically significant national bias, awarding inflated scores to compatriot athletes. However, this bias did not intensify when both athletes' nationalities were represented on the judging panel, nor did it increase as the competition advanced from the round-robin stage to the final battles. These patterns suggest that national bias was not applied strategically, and its occurrence in only a subset of judges indicates that partiality is unlikely to be a pervasive issue across the judging body. Nevertheless, individual differences in the susceptibility to bias or the strategic use of scoring margins may exist and warrant closer examination.

Overall, the findings underscore the need for continued investigation into the scoring systems used in aesthetic sports and the judging behavior in the context of sports competitions. As a relatively young discipline that continues to gain prominence on the global stage and to experience development of its scoring rules and conceptions of performance quality, Breaking provides a unique context for studying the development of a highly distinctive scoring and competition format in which two athletes compete directly against one another in an aesthetic sports setting.